\documentclass[12pt]{article}

\usepackage{graphicx}
\usepackage{epsfig}
\usepackage{amsfonts}
\usepackage{amssymb}
\usepackage{amsmath,amssymb}

\newcommand{\ee}{\end{equation}}
\newcommand{\eea}{\end{eqnarray}}
\newcommand{\be}{\begin{equation}}
\newcommand{\bea}{\begin{eqnarray}}

\begin{document}

\title{  Anti-de-Sitter regular electric multipoles:   \\
towards Einstein-Maxwell-AdS solitons}

\author{
{\large Carlos Herdeiro}\footnote{herdeiro@ua.pt} \
and
{\large Eugen Radu}\footnote{eugenradu@ua.pt}
\\ 
\\
{\small Departamento de F\'\i sica da Universidade de Aveiro and CIDMA} \\ 
{\small   Campus de Santiago, 3810-183 Aveiro, Portugal}
}
\date{May 2015}
\maketitle

\begin{abstract}  
We discuss electrostatics in Anti-de-Sitter ($AdS$) spacetime, in global coordinates. We observe that the multipolar expansion has two crucial differences to that in Minkowski spacetime. First, there are everywhere regular solutions, with finite energy, for every multipole moment \textit{except} for the monopole. Second, all multipole moments decay with the same inverse power of the areal radius, $1/r$, as spatial infinity is approached. The first observation suggests there may be regular, self-gravitating, Einstein-Maxwell solitons in $AdS$ spacetime. 
The second observation, renders a Lichnerowicz-type no-soliton theorem inapplicable. Consequently, we suggest Einstein-Maxwell solitons exist in $AdS$, and we support this claim by computing the first order metric perturbations sourced by test electric field multipoles, which are obtained analytically in closed form. 
 
\end{abstract}

\section{Introduction}
Anti-de-Sitter ($AdS$) gravity has been under intense scrutiny, motivated, in particular, by the gauge-gravity duality~\cite{Maldacena:1997re}. A central question one may ask, as for asymptotically flat gravity, is under which circumstances everywhere regular gravitating solitons, admitting a global timelike Killing vector field exist. Whereas such solutions are known for non-linear matter sources  -- for instance Yang-Mills fields~\cite{BoutalebJoutei:1979va,Bjoraker:2000qd,Bjoraker:1999yd} -- Lichnerowicz-type arguments have been put forward to preclude the existence of such stationary configurations in Einstein's theory with a negative cosmological constant vacuum~\cite{Boucher:1983cv} or electro-vacuum~\cite{Shiromizu:2012hb}. The purpose of this paper is provide evidence that, in the static electro-vacuum case, such arguments do not apply, and, moreover, a linear field analysis suggests the existence of Einstein-Maxwell-$AdS$ gravitating solitons.

From elementary electromagnetism in Minkowski spacetime, any static charge distribution can be described by a multipolar expansion for the electrostatic potential, which in standard spherical coordinates takes the form $\phi(r,\theta,\phi)=\sum_{\ell,m} R_\ell(r) Y_\ell^m(\theta,\phi)$, where $Y_\ell^m(\theta,\phi)$ are spherical harmonics. It is well known that $R_\ell(r)=c_1 r^\ell+c_2/r^{\ell+1}$. Thus, any non-trivial solution diverges either at the origin or at infinity. Moreover, the total energy associated to any such multipolar field is infinite. Finite energy configurations can, however, be obtained, by confining the electric field to be inside a box -- say, spherical and of radius $r_B$ -- and putting $c_2=0$. Then, a unique electric field with total finite energy is obtained by specifying the boundary data $R_\ell(r_B)$, which determines the set of non-trivial multipoles at the boundary. Thus, ``boxing" Minkowski spacetime allows for everywhere regular electric multipoles with a finite total energy inside the box.

In gravitational physics, $AdS$ is a natural ``box", in view of its conformal timelike boundary. Thus, it seems worth investigating electrostatics in $AdS$ and inquiring if everywhere regular finite energy multipoles exist. Even though this question seems natural -- even obvious -- to the best of our knowledge it has not been discussed in the literature. The first goal of this paper is to show that global $AdS$, unlike global Minkowski, admits everywhere regular, finite energy electric fields for \textit{all multipoles except the monopole}. Moreover, as for the boxed-Minkowski region described above, a unique regular electric field configuration is determined by specifying boundary data which identifies the excited multipoles at the boundary.

 A second difference with the Minkowski case is that as one approaches the boundary, all of the above regular electric multipoles decay with the same inverse power of the areal radius, $1/r$. The second goal of this paper is to show that this behaviour 
 invalidates a Lichnerowicz-type no-soliton theorem for the Einstein-Maxwell system in 
 $AdS$~\cite{Shiromizu:2012hb}. Thus, there appears to be no obstruction to the promotion of the regular electric fields in $AdS$ to fully non-linear gravitating solutions within Einstein-Maxwell-$AdS$ theory. In fact, we shall provide evidence for the existence of such solitonic solutions by computing the first order perturbation of $AdS$ sourced by the regular test electric multipole fields mentioned above. The fully non-linear configurations, 
 whose existence we anticipate, will yield Einstein-Maxwell-$AdS$ solitons. 

This paper is organized as follows. In Section~\ref{sec_2} we present  Einstein-Maxwell-$AdS$ gravity and we briefly review the argument for the non-existence of solitons in this theory. In Section~\ref{sec_3} we discuss electrostatics in $AdS$, presenting the everywhere regular electric multipoles, their asymptotic behaviour, their energy density and their total energy. Then, in Section~\ref{sec_4} we discuss the backreaction on $AdS$ of these electric multipoles within first order perturbation theory, obtaining explicit results for the dipole. In Section~\ref{sec_5} we draw our conclusions and further remarks.

\section{Solitons in Einstein-Maxwell-$AdS$ gravity?}
\label{sec_2}
We are interested in Einstein-Maxwell theory in the presence of a negative cosmological constant (herein dubbed \textit{Einstein-Maxwell-AdS gravity}):
\begin{eqnarray}
\mathcal{S} =\int d^4 x\sqrt{-g}\left\{\frac{1}{16\pi G}\left(R-2\Lambda\right)
 -\frac{1}{4}F_{\mu \nu}F^{\mu\nu}\right\} \ ,
 \label{EMAdS}
\end{eqnarray}
where $F=dA$ is the $U(1)$ field strength and $\Lambda\equiv -3/L^2<0$ is the cosmological constant, with $L$ the $AdS$ ``radius". Varying the action one obtains the Einstein equations:
 \begin{eqnarray} 
E_{\mu\nu}\equiv R_{\mu\nu}-\frac{1}{2}R g_{\mu\nu}+\Lambda g_{\mu\nu}-8\pi G T_{\mu\nu}=0 \ ,
\label{eineq}
\end{eqnarray} 
and the Maxwell equations
\begin{eqnarray}
d\star F=0 \ .
\label{me}
\end{eqnarray}
The electromagnetic energy-momentum tensor is
\begin{eqnarray}
T_{\mu\nu}=F_{\mu \alpha}F_{\nu\beta}g^{\alpha \beta}-\frac{1}{4}g_{\mu\nu}F^2\ .
\label{emtensor}
\end{eqnarray}
The maximally symmetric solution of this theory is $AdS$ spacetime (with $F=0$),
 which in global coordinates reads
\begin{eqnarray}
\label{AdS}
ds^2=-N(r)dt^2+\frac{dr^2}{N(r)}+r^2(d \theta^2+\sin^2\theta d\varphi^2)
,~~{\rm where}~~N(r)=1 +\frac{r^2}{L^2} \ .
\end{eqnarray} 
Boucher, Gibbons and Horowitz~\cite{Boucher:1983cv}  showed that, in the absence of the Maxwell term, \eqref{AdS} is the only strictly stationary -- $i.e$ admitting an everywhere timelike Killing vector field $k^\mu$ (which excludes black hole regions) -- and asymptotically $AdS$ solution of this theory. 
Thus, there are no regular, finite-energy, time-independent solutions -- or \textit{gravitating solitons} -- in vacuum $AdS$ 
gravity\footnote{
Ref.~\cite{Horowitz:2014hja} provides numerical evidence for the existence of vacuum, albeit not stationary, $AdS$ geons. 
}. 
Let us briefly review the argument including the Maxwell field~\cite{Shiromizu:2012hb}, 
but restricting to the static case and a purely electric Maxwell field, of interest herein.  
In \cite{Shiromizu:2012hb}, the following identity was shown, which is implied by the Einstein-Maxwell equations:
\begin{eqnarray}
\label{relS}
 \nabla_{\mu} \left(\frac{\nabla^\mu V^2}{V^2}-\tau^\mu+W^\mu \right)=0 \ ,
\end{eqnarray} 
where 
$V^2=-k_\mu k^\mu$,
$\nabla_\mu\tau^\mu=-2\Lambda$
and 
$W^\mu=-8\pi G \frac{\phi E^{\mu} }{V^2}$; also,  $E_{\mu}=k^{\nu}F_{\nu\mu}=\nabla_\mu \phi$,
such that when choosing a time coordinate $t$ adapted to the Killing field, $k=\partial/\partial t$, $\phi$ becomes the electrostatic potential. Upon integration over a spacelike hypersurface, 
equation (\ref{relS}) is converted into a surface integral at infinity by Stokes theorem. Observe that one is assuming the absence of an event horizon, such that there is no contribution
from an inner boundary term.  
If the $W^r$ term vanishes faster than $1/r^2$, as implicitly assumed in \cite{Shiromizu:2012hb}, then it does not contribute to the boundary term. It follows that we are left with the first (purely geometric) two terms in  (\ref{relS}), which yield a vanishing total mass for spacetime, $M=0$ 
\cite{Boucher:1983cv,Shiromizu:2012hb}. 
Then, by the positive energy theorem, the spacetime must be $AdS$.

The assumption made on the asymptotic decay of $W^r \sim \phi F^{rt}$ is
justified for Maxwell multipoles in asymptotically flat spacetime.
As shown in the next section, however, the Maxwell equations in a fixed AdS background
possess everywhere regular configurations with  $W^r \sim 1/r^2$
in the far field, where $r$ is the areal radius, whose existence thus invalidates the above no-soliton theorem.

\section{Electrostatics on $AdS$: regular electric multipoles}
\label{sec_3}
We linearize the model (\ref{EMAdS}) around  an empty $AdS$ background. 
Thus we solve the (test) Maxwell equations (\ref{me}) on the geometry (\ref{AdS}). 
Here we shall focus on axially symmetric electric potentials:
\begin{eqnarray}
A=\phi_{\ell}(r,\theta)\equiv R_\ell(r) \mathcal{P}_\ell(\cos \theta) dt\ , 
\label{ep}
\end{eqnarray}
where 
 $\mathcal{P}_\ell$ is a Legendre polynomial of degree $\ell$, with $\ell\in \mathbb{N}_0$ defining the multipolar structure. Then Maxwell's equations reduce to the radial equation
\begin{eqnarray}
\label{eq}
\frac{d}{dr} \left(r^2\frac{dR_\ell(r)}{dr}\right)=\frac{\ell(\ell+1)}{N(r)}R_\ell\ .
\label{radial}
\end{eqnarray}
In the Minkowski spacetime limit,  $L\rightarrow \infty$, it is well known from elementary electrostatics that the solution of (\ref{radial}) is of the form $R_\ell(r)=c_1 r^\ell+c_2/r^{\ell+1}$. Thus, any non-trivial solution diverges either at the origin or at infinity. By contrast, in $AdS$, non-trivial solutions exist which are \textit{regular everywhere}. 
To see this, take the ansatz
\begin{equation}
R_\ell(r)=\left(\frac{r}{L}\right)^\ell f_\ell(r),
\end{equation} 
such that
the radial equation (\ref{radial}) becomes the hypergeometric equation (with $x\equiv-\frac{r^2}{L^2}$)
\begin{equation}
x(1-x)\frac{d^2f_\ell(x)}{dx^2}+(c-(a+b+1)x)\frac{df_\ell(x)}{dx}-ab f_\ell(x)=0 \ ,
\end{equation}
where 
$a=(\ell+1)/2$, 
$b=\ell/2$ and 
$c=\ell+3/2$. 
Since, for any $\ell$, either $a$ or $b$ is an integer, this hypergeometric equation is degenerate, \textit{i.e.} one of the independent solutions is a finite polynomial. For $\ell\ge 1$, the everywhere regular solution, in particular at $r=0$, is, in terms of the hypergeometric function ${}_2F_1$:
\begin{eqnarray}
R_\ell(r)=
\displaystyle{\frac{\Gamma(\frac{1+\ell}{2})\Gamma(\frac{3+\ell}{2})}{\sqrt{\pi}\Gamma(\frac{3}{2}+\ell)}
 \frac{r^{\ell}}{L^\ell}~{}_2F_1\left(\frac{1+\ell}{2} , \frac{ \ell}{2}   ;  \frac{3}{2}+\ell  ;  - \frac{r^2}{L^2}\right)} \ ,
\end{eqnarray} 
where we have normalized it such that $R_\ell(r)\to 1$ asymptotically.

The case $\ell=0$ is a trivial solution: $R_0(r)=1$. 
For $\ell\ge 1$ the solutions are, however, non-trivial. 
Simplified expressions for $\ell=1,2,3$ are:
\begin{eqnarray}
\nonumber
&&
R_1(r)=-\frac{2}{\pi}\left[\frac{L}{r}-\left(1+\frac{L^2}{r^2}\right)\arctan\left(\frac{r}{L}\right)\right] \ ,
\\
\label{r1}
&&
R_2(r)=1+\frac{3L^2}{2r^2}-\frac{3L}{2r}\left(1+\frac{L^2}{r^2}\right)\arctan\left(\frac{r}{L}\right) \ ,
 \\
&&
\nonumber
R_3(r)=-\frac{2 }{\pi}
\left[
\frac{L}{3r}\left(13+\frac{15L^2}{r^2}\right)-\left(1+\frac{L^2}{r^2}\right)
\left(1+\frac{5L^2}{r^2} \right)\arctan\left(\frac{r}{L}\right) 
\right ]~.
 \end{eqnarray} 
\begin{figure}[h!]
\begin{center}
\includegraphics[width=0.7\textwidth]{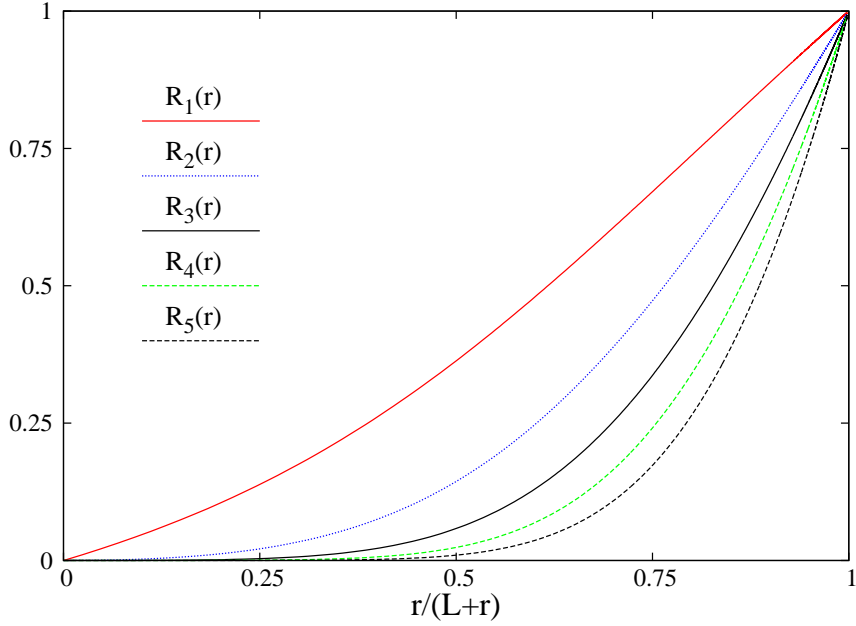}
\caption{The radial function $R_{\ell}$ for the regular electric multipoles in $AdS$ with $\ell=1,2,3,4,5$ and with the normalization $R_\ell \to 1$ asymptotically, in terms of a compactified coordinate $r/(L+r)$. Here and in  Figures 2, 3 we take an $AdS$ radius $L=1$. Observe that the $R_\ell(r)$'s are nodeless functions. This is a general feature.}
\label{R}
\end{center}
\end{figure} 

We notice also the existence of the recurrence relation
\begin{equation}
R_{\ell+2}(r)=R_\ell(r)-\alpha_\ell \frac{L}{r}R_{\ell+1}(r) \ ,
\end{equation}
where $\alpha_\ell$ are positive constants:
\begin{equation}
\alpha_0=\frac{3\pi}{4} \ , \qquad \alpha_\ell=\frac{(3+2\ell)}{\ell}\frac{\Gamma(\frac{\ell+1}{2})\Gamma\left(\frac{3+\ell}{2}\right)}{ \Gamma(\ell/2)\Gamma(2+\ell/2)} \ , \ \ \ell\ge 1 \ .
\end{equation}
In Figure~\ref{R} we exhibit $R_{\ell}$, for $\ell=1,2,3,4,5$.

As one would expect, at the origin, the $AdS$ regular multipoles approach the behaviour of the  Minkowski mulipoles which are regular therein. Indeed, as $r\to 0$, the solutions behave as
 \begin{eqnarray}
R_\ell(r)=c_0^{(\ell)} \left(\frac{r}{L}\right)^\ell+\dots \ , \qquad {\rm where} 
\qquad 
 c_0^{(\ell)}=  \frac{ \Gamma\left(\frac{1+\ell}{2}\right) \Gamma\left(\frac{3+\ell}{2}\right)}{\sqrt{\pi} \Gamma\left(1+\frac{\ell}{2}\right)} \ .
\end{eqnarray} 
By contrast, asymptotically, the regular $AdS$ multipoles have a remarkably different behaviour from that of the Minkowski multipoles which are regular at infinity. Indeed, as $r\rightarrow \infty$, the solutions become 
\begin{eqnarray}
\label{as1}
R_\ell(r)=1-c_1^{(\ell)}\frac{L}{r}+\dots\ , \qquad {\rm where} \qquad
%
c_1^{(\ell)}= \frac{2 \Gamma\left(\frac{1+\ell}{2}\right) \Gamma\left(\frac{3+\ell}{2}\right)}{ \Gamma\left(1+\frac{\ell}{2}\right) \Gamma\left(\frac{\ell}{2}\right)}\ .
\end{eqnarray}
%
In particular observe that all multipoles fall-off with the same $1/r$ power, where $r$ is the areal radius, $cf.$ eq. (\ref{AdS}).

\subsection{Energy density and total energy}

\begin{figure}[h!]
\begin{center}
\includegraphics[width=0.7\textwidth]{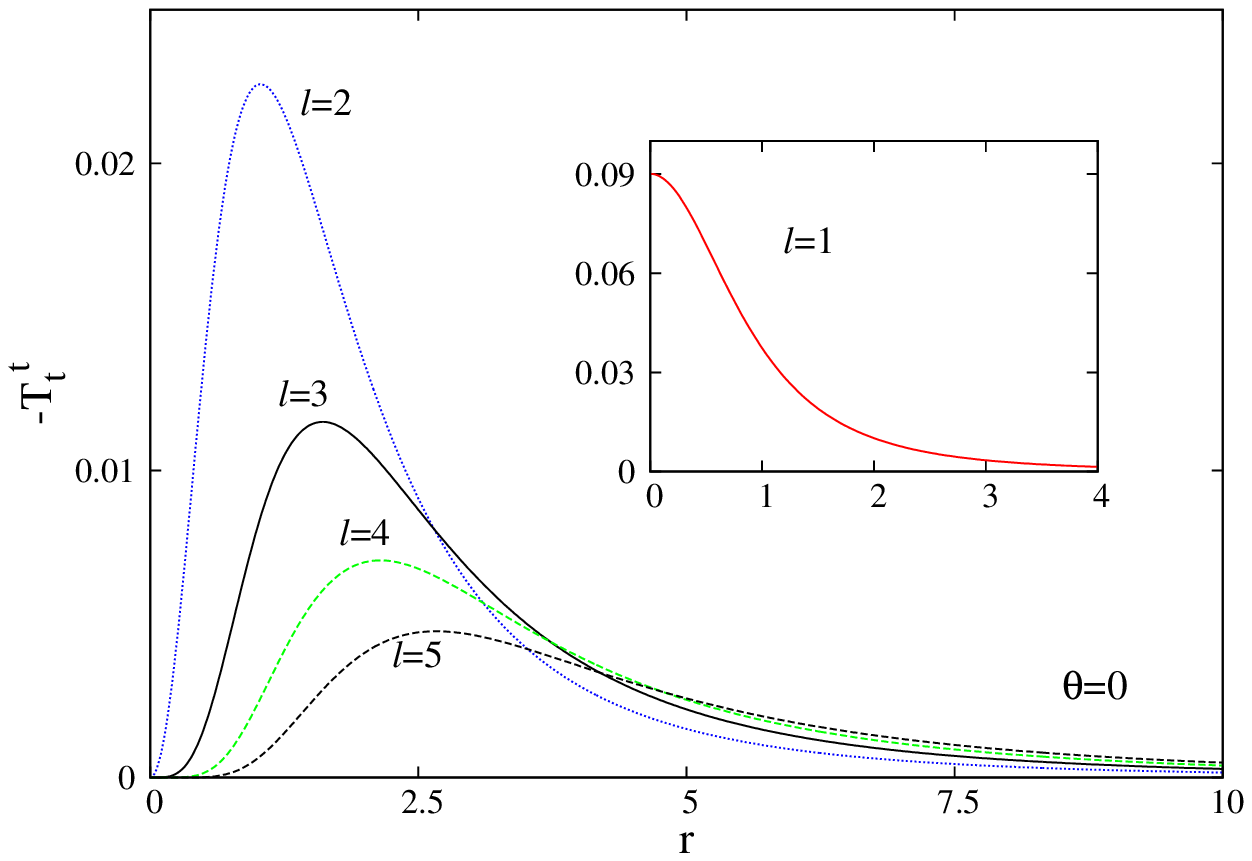}
\includegraphics[width=0.7\textwidth]{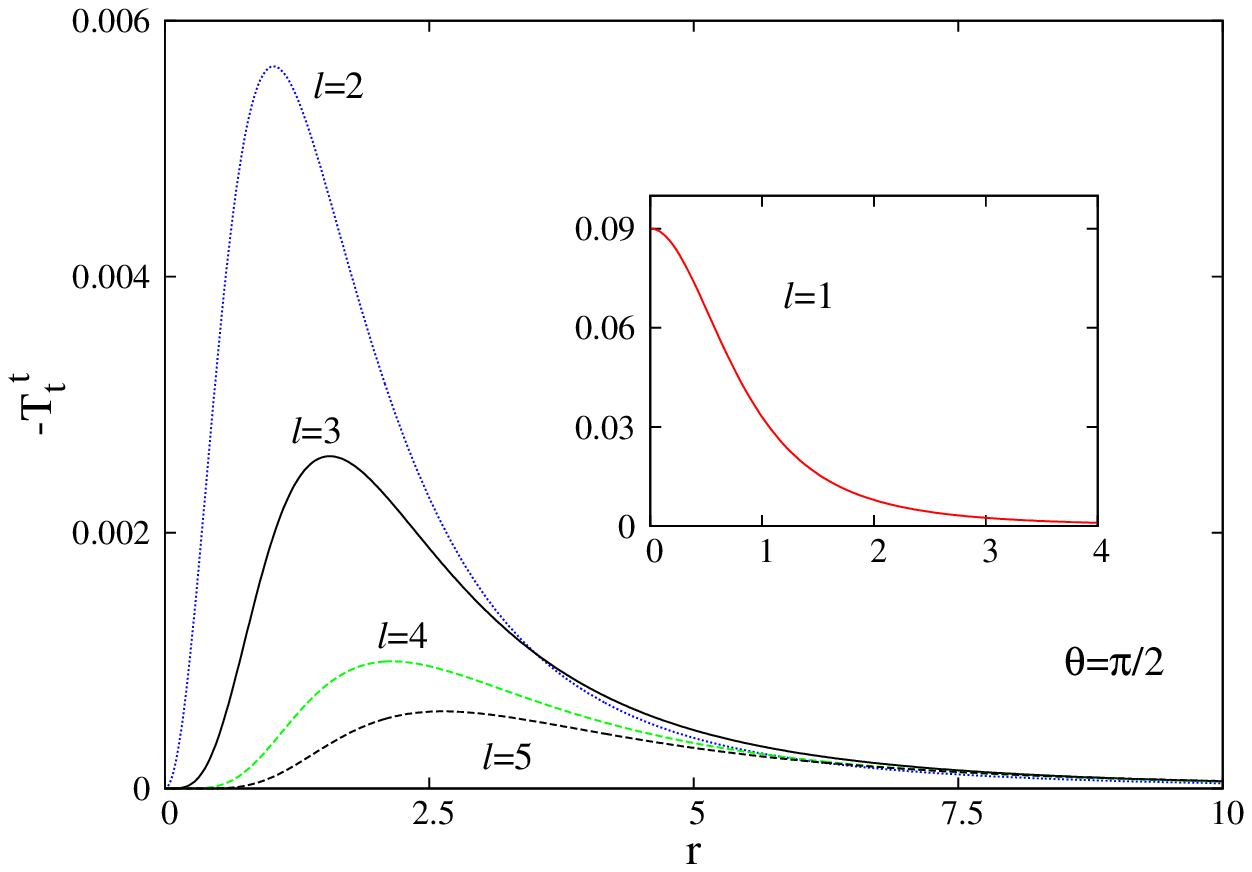}
\caption{Energy density $\rho$, of the $AdS$ regular electric multipoles $\phi_{\ell}$, for $\ell=1,2,3,4$, with the normalization $R_\ell(r)\to 1$ asymptotically, in terms of the radial coordinate $r$. The top (bottom) panel correspond to sections with $\theta=0$ ($\theta=\pi/2$).}
\label{E}
\end{center}
\end{figure}

The energy density of a given electromagnetic configuration, $\rho$, as measured by a static observer with 4-velocity $U^\mu\propto \delta^\mu_t$, is $\rho=-T_t^t$. For the above regular electric multipoles, $\phi_{\ell}$ given by eq. (\ref{ep}), $\rho$ is  finite everywhere.
 This is illustrated in Figure~\ref{E}.

%

As can be seen there, for $\ell=1$, 
$\rho$ is maximal at the origin and monotonically decreasing with $r$.
 For $\ell>1$, $\rho=0$ at the origin, attains a maximum at some radius $r$ which increases with $\ell$ and then decreases monotonically with $r$. This is an expected behaviour in terms of the angular momentum harmonic index.

At infinity, $\rho$ decays as $1/r^4$. Consequently,  the  total energy of these solutions
\begin{eqnarray}
E=-\int d^3 x  \sqrt{-g}\, T_t^t=-2\pi \int_0^\infty dr  \int_0^\pi d\theta~r^2  \sin \theta T_t^t\ ,
\end{eqnarray} 
is finite.  Indeed, with the chosen normalization, 
the total energy for the first few multipoles $E_{\ell}$ is
\begin{eqnarray}
\label{Ei}
E_{1}=\frac{8 L}{3}\ , \ \ \ E_{2}=\frac{3\pi^2 L}{10}\ , \ \ \ E_{3}=\frac{64 L}{21}~.
\end{eqnarray} 

The total energy can actually be expressed as a surface integral. Noticing that 
\begin{eqnarray}
E = -\pi \lim_{r\to \infty}\int_0^\pi  r^2 \sin \theta A_t F^{rt} d\theta \ ,
\end{eqnarray} 
we obtain, for a given mode $\ell$, 
 \begin{eqnarray}
E_{\ell}=  \pi  \lim_{r\to \infty}r^2 R'_\ell(r)\int_0^\pi d\theta \sin \theta L^2_\ell(\cos\theta)\ .
\end{eqnarray}
From (\ref{as1})  together with the properties of the Legendre polynomial one finds a general formula for the total energy
\begin{equation}
\label{energy}
E_{\ell}=\frac{4\pi}{2\ell+1}\frac{ \Gamma(\frac{1+\ell}{2}) \Gamma(\frac{3+\ell}{2})}{ \Gamma(1+\frac{\ell}{2}) \Gamma(\frac{\ell}{2})}  L \ .
\end{equation}

We observe that the regular electric multipoles $\phi_{\ell}$ satisfy the virial identity
   \begin{eqnarray}
 \int_0^\infty r^2 dr \int_0^\pi \sin \theta
 \left (
 \left[\phi_{\ell}\right]_{,r}^2+\frac{1-\frac{r^2}{L^2}}{N^2(r)r^2} \left[\phi_{\ell}\right]_{,\theta}^2
\right )=0 \ .
\end{eqnarray}
For a vanishing cosmological constant, $N(r)=1$. 
Then both terms in the integrand are positive definite and the electric potential must be trivial, 
for the equality to hold. 
This makes explicit that the $AdS$ geometry supplies the attractive force needed 
to balance the repulsive force of the gauge interactions. 
Also, it is clear that the configurations are supported by the nontrivial angular dependence of the electrostatic potential.

\section{First order backreaction of the electric multipoles}
\label{sec_4}
The existence of a solution in the probe limit is usually taken as a strong indication that the full 
system possess nontrivial gravitating configurations. As an example with gauge fields, let us mention the case of the Yang-Mills (test field) 
static soliton in a fixed $AdS$ background discovered in \cite{BoutalebJoutei:1979va}. 
As found in \cite{Bjoraker:2000qd,Bjoraker:1999yd}, 
this configuration can be promoted to a gravitating soliton of the full theory when including its backreaction.

In the same spirit, our goal here is to consider how the $AdS$ geometry (\ref{AdS}) is deformed if one takes into account the backreaction of the regular electric multipoles discussed in Section~\ref{sec_3}.
For $\Lambda=0$, the axially symmetric  Einstein-Maxwell solutions can be constructed in closed form,
by using powerful analytical techniques. Unfortunately, such methods do not extend to the $AdS$ case.
In this paper  we solve the Einstein equations (\ref{eineq}) in first order perturbation theory, sourced by the energy-momentum tensor~(\ref{emtensor}) of the regular electric multipoles. We take the Maxwell 4-potential to have the generic form:
 \begin{eqnarray} 
A=V(r,\theta)dt \ ,
\end{eqnarray} 
and we consider a metric ansatz using the gauge choice
\begin{eqnarray} 
ds^2=-e^{2\nu(r,\theta)}dt^2+e^{2\mu(r,\theta)}dr^2+e^{2\psi(r,\theta)}r^2(d\theta^2+\sin^2\theta d\varphi^2) \ .
\label{metrica}
\end{eqnarray} 
This depends on three arbitrary functions of $(r,\theta)$ which covers the general case of an axisymmetric, \textit{static} geometry.

To setup the perturbative expansion, we introduce a perturbative parameter $\alpha$, which multiplies the regular electric multipoles computed in Section~\ref{sec_3}, $\phi_{\ell,m}$. 
Thus $\alpha$ can be identified with the  magnitude of the electric potential at infinity
(here we set $4\pi G=1$). 
Then, we expand the three metric functions plus the electric potential as:
\begin{eqnarray} 
&&
\nu(r,\theta)=\nu^{(0)}(r)+\alpha^2\nu^{(2)}(r,\theta)+\dots\ ,
\nonumber
\\
&&
\label{mf1}
\mu(r,\theta)=\mu^{(0)}(r)+\alpha^2\mu^{(2)}(r,\theta)+\dots\ ,
\\
&&
\psi(r,\theta)=\alpha^2\psi^{(2)}(r,\theta)+\dots\ ,
\nonumber
\\
\nonumber
&&
V(r,\theta)=\alpha V^{(1)}+\dots\ ,
\end{eqnarray} 
where
\begin{eqnarray} 
\nu^{(0)}(r)\equiv \frac{1}{2}\log\left(1+\frac{r^2}{L^2}\right) \equiv  -\mu^{(0)}(r)\ \ , \qquad V^{(1)}\equiv \phi_{\ell}(r,\theta)=R_\ell(r)\mathcal{P}_\ell(\cos\theta) \ .
\label{order0}
\end{eqnarray} 
Observe that the bracketed superscript denotes the order in $\alpha$ of the corresponding term. Our goal is to compute the metric perturbation to $\mathcal{O}(\alpha^2)$ using the  $\mathcal{O}(\alpha)$ Maxwell field.


The derivation of a general solutions starts with the observation that
the Einstein equation
\begin{eqnarray} 
E_\theta^\theta-E_\varphi^\varphi=0 \ ,
\end{eqnarray} 
implies
\begin{eqnarray} 
\frac{d^2U}{d\theta^2}-\cot \theta \frac{d U}{d\theta }=\frac{2}{N}\left[V^{(1)}_{,\theta}\right]^2 \ ,
\end{eqnarray}  
where
\begin{eqnarray} 
U(r,\theta)\equiv \nu^{(2)}+\mu^{(2)} \ .
\end{eqnarray} 
Using the second equation in (\ref{order0}),  the function $U$ can be expressed as
\begin{eqnarray} 
U(r,\theta)=\frac{[R_\ell(r)]^2}{N(r)}Q_\ell(\cos\theta) \ ,
\end{eqnarray}
where the function $Q_\ell(x)$ is a solution of the equation
\begin{eqnarray} 
\label{t1}
\frac{d^2Q_\ell}{dx^2}=2\left(\frac{d\mathcal{P}_\ell(x)}{dx}\right)^2 \ .
\end{eqnarray}
Thus, $Q_\ell$ can be, itself, expanded in Legendre polynomials:
%
\begin{eqnarray} 
Q_\ell(\cos\theta)=\sum_{j=0}^\ell a_j \mathcal{P}_{2j}(\cos \theta)\ ,
\end{eqnarray}
for fixed coefficients $a_j$.

Let us illustrate the method by considering the lowest mode $\ell=1$. 
Thus, for the remaining of this section we compute the $AdS$ perturbations 
by a regular electric dipole field. For this case:
\begin{eqnarray} 
Q_1(\cos\theta)=\frac{1}{3}+\frac{2}{3}\mathcal{P}_{2}(\cos\theta)\ .
\end{eqnarray}
Assembling the results, and recalling the explicit form of $R_1(r)$, $cf.$ eq. (\ref{r1}), we obtain that 
\begin{equation}
 \nu^{(2)}=-\mu^{(2)} +  \frac{4L^2}{\pi^2 r^2N(r)}\left(-1+\frac{L}{r}N(r)\arctan\frac{r}{L} \right)^2\left[\frac{1}{3}+\frac{2}{3}\mathcal{P}_{2}(\cos\theta)\right] \ .
 \label{nu2}
 \end{equation}
We now regard $\nu^{(2)}$ as given by this relation: thus we focus on the computation of $\mu^{(2)}$ and $\psi^{(2)}$.

For the next step, we consider a consistent expansion of the angular dependence of the two remaining metric functions in terms of Legendre polynomials: 
\begin{eqnarray} 
\label{expansion}
&&
\mu^{(2)}(r,\theta)=\mu^{(2,0)}(r)+\mu^{(2,1)}(r)\mathcal{P}_1(\cos\theta)+\mu^{(2,2)}(r)\mathcal{P}_2(\cos\theta) \ ,
\\
\nonumber
&&
\psi^{(2)}(r,\theta)=\psi^{(2,0)}(r)+\psi^{(2,1)}(r)\mathcal{P}_1(\cos\theta)+\psi^{(2,2)}(r)\mathcal{P}_2(\cos\theta) \ .
\end{eqnarray} 
Observe that there is now a second number in the superscript, referring to the order of the expansion in Legendre polynomials. 
One can show that the first order terms in the angular expansion vanish:
\begin{eqnarray} 
\mu^{(2,1)}=0= 
\psi^{(2,1)} \ .
\end{eqnarray} 
Then, a direct computation leads to the remaining terms of the solution. The zeroth order terms in the angular expansion 
(\ref{expansion})
are:
\begin{eqnarray} 
\mu^{(2,0)} = \frac{\psi^{(2,0)}}{N}+
\frac{1}{r N} 
\int dr 
\left[
\frac{2r^3}{L^2}\frac{d\psi^{(2,0)}}{dr}
+\frac{1}{6}
\bigg( 2R_1+r R_1' \bigg)^2- \bigg(\frac{r^2}{3L^2N}+1  \bigg)R_1^2
\right] \ ,
\end{eqnarray}
where $\psi^{(2,0)}$
is found by integrating the equation
\begin{eqnarray} 
\frac{ d}{dr}(r\psi^{(2,0)})= -\frac{1}{2}
+\frac{2}{\pi^2}
\left(
\frac{L^2}{r^2}+\frac{2L}{r}\left[1-\frac{L^2}{r^2}\right]\arctan\left[\frac{r}{L}\right]
+(1+\frac{2L^2}{3r^2}+\frac{L^4}{r^4})\arctan^2\left[\frac{r}{L}\right]
\right) .
\label{psi20}
\end{eqnarray}
 Although 
 $\mu^{(2,0)}$, 
 $\psi^{(2,0)}$
 do not appear to have a simple expression in terms of elementary functions,
 one can evaluate them numerically.
 Also, an approximate expression can be found for $r\to 0$,
 with
\begin{eqnarray} 
 \mu^{(2,0)}=\frac{16}{3\pi^2}-\frac{1}{2}-\frac{27 \pi^2-272}{54 \pi^2 \pi^2}\frac{r^2}{L^2}+\mathcal{O}(r^4)\ ,
~~ \psi^{(2,0)}=\frac{16}{3\pi^2}-\frac{1}{2}-\frac{16}{675\pi^2}\frac{r^4}{L^4}+\mathcal{O}(r^6)\ ,
\end{eqnarray} 
and for large $r$
\begin{eqnarray} 
 \mu^{(2,0)}=-p_1\frac{L}{r}+\frac{2}{3}\frac{L^2}{r^2}+\mathcal{O}\left(\frac{1}{r^3}\right)\ ,
~~ \psi^{(2,0)}=p_1\frac{L}{r}-\frac{1}{3}\frac{L^2}{r^2}+\mathcal{O}\left(\frac{1}{r^3}\right) \ ,
\end{eqnarray} 
with $p_1\simeq 0.164 .$h

 The second order terms in the  expansion (\ref{expansion}), on the other hand, can be written explicitly. Here and in the expressions of \
$\mu^{(2,0)}$ 
and
$\psi^{(2,0)}$ 
 a number of integration constant were chosen such that the solutions are regular at $r=0$
and vanish at infinity. Then: 
\begin{eqnarray} 
 \label{mu22}
\mu^{(2,2)} =  
&&
\left(
5+\frac{3L^2}{r^2}+\frac{8}{3\pi^2}
\left[5+\frac{7L^2}{r^2}\right]
\right)
\frac{1}{8N}
+\frac{4 N}{3\pi^2}\frac{L^4}{r^4}\arctan^2\left[\frac{r}{L}\right]
\\
\nonumber
&&
-\frac{3L^3}{8r^3}
\left(
N+\frac{8}{9\pi^2}\frac{11+\frac{14r^2}{L^2}-\frac{5r^4}{L^4}}{N}
\right)
\arctan\left[\frac{r}{L}\right]~,
\end{eqnarray}
and
\begin{eqnarray} 
 \label{psi22}
\psi^{(2,2)} = 
&&
-\frac{1}{2}+\left(\frac{1}{3\pi^2}+\frac{3}{8} \right)\frac{L^2}{r^2}
+\frac{L}{r}\left(\frac{3}{8}\left[1-\frac{L^2}{r^2}\right]+\frac{7+\frac{L^2}{r^2}}{3\pi^2}\right)\arctan\left[\frac{r}{L}\right]
\\
\nonumber
&&
+\frac{2}{3 \pi^2}
\left(
3-\frac{4L^2}{r^2}-\frac{L^4}{r^4}
\right)
\arctan^2\left[\frac{r}{L}\right]~.
\end{eqnarray}
These functions behave as
\begin{eqnarray} 
 \label{exp2-0}
 &&
 \mu^{(2,2)}=\left(\frac{368}{135\pi^2}-\frac{1}{5}\right)\frac{r^2}{L^2}+\left(\frac{8}{35}-\frac{3568}{945\pi^2}\right)\frac{r^4}{L^4}+\mathcal{O}(r^6)\ ,~~
 \\
 &&
 \nonumber
 \psi^{(2,2)}=\left(\frac{368}{135\pi^2}-\frac{1}{5}\right)\frac{r^2}{L^2}+\left(\frac{9}{70}-\frac{1888}{945\pi^2}\right)\frac{r^4}{L^4}+\mathcal{O}(r^6)\ ,~~
 \end{eqnarray}
for $r\to 0$ and
 \begin{eqnarray} 
 \label{exp2-inf}
 \mu^{(2,2)}=\left(\frac{5}{6 \pi}-\frac{3\pi}{16}\right)\frac{L}{r}+\frac{4}{3}\frac{L^2}{r^2}+\mathcal{O}\left(\frac{1}{r^3}\right) \ ,~~
 \psi^{(2,2)}=\left(-\frac{5}{6 \pi}+\frac{3\pi}{16}\right)\frac{L}{r}-\frac{2}{3}\frac{L^2}{r^2}+\mathcal{O}\left(\frac{1}{r^3}\right)\ ,~~
 \end{eqnarray}
for large-$r$.

 \begin{figure}[h!]
\begin{center}
\includegraphics[width=0.7\textwidth]{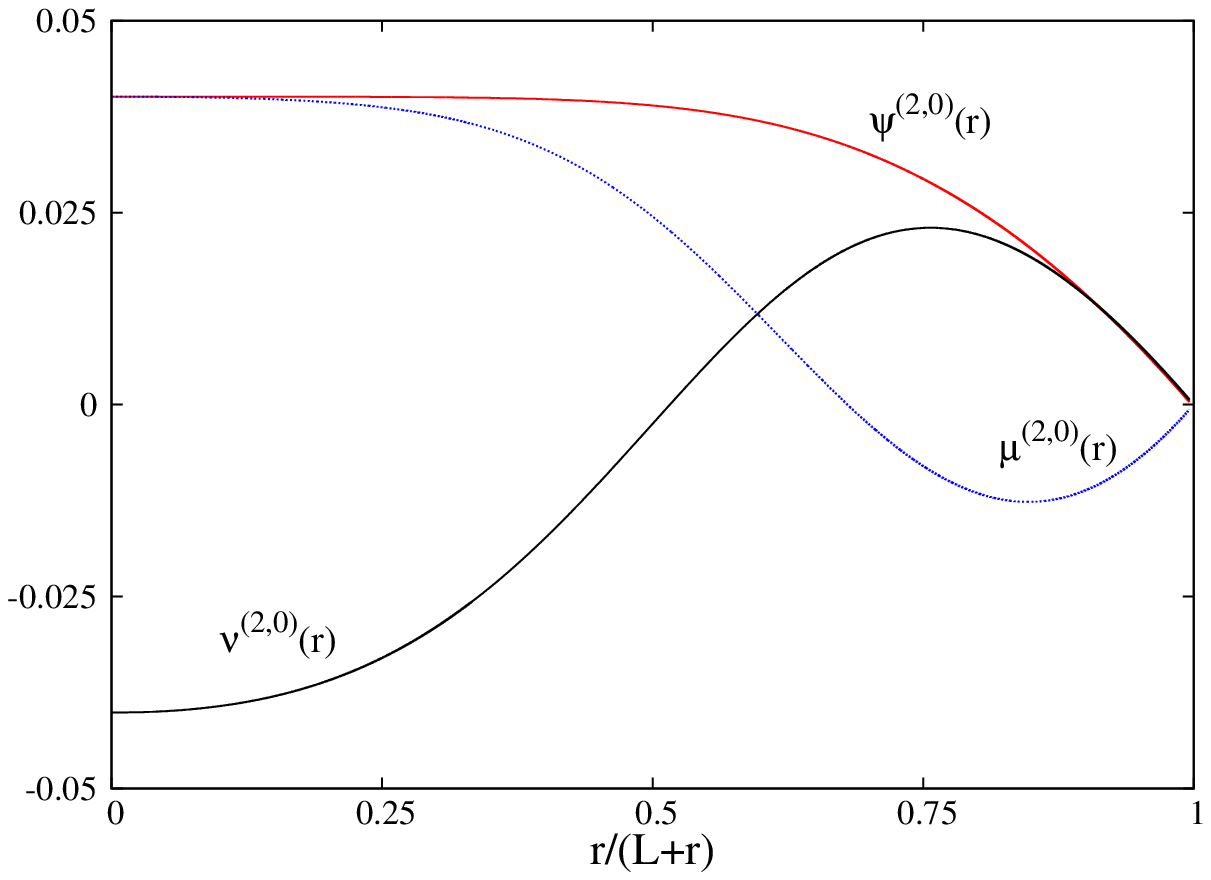}
\includegraphics[width=0.7\textwidth]{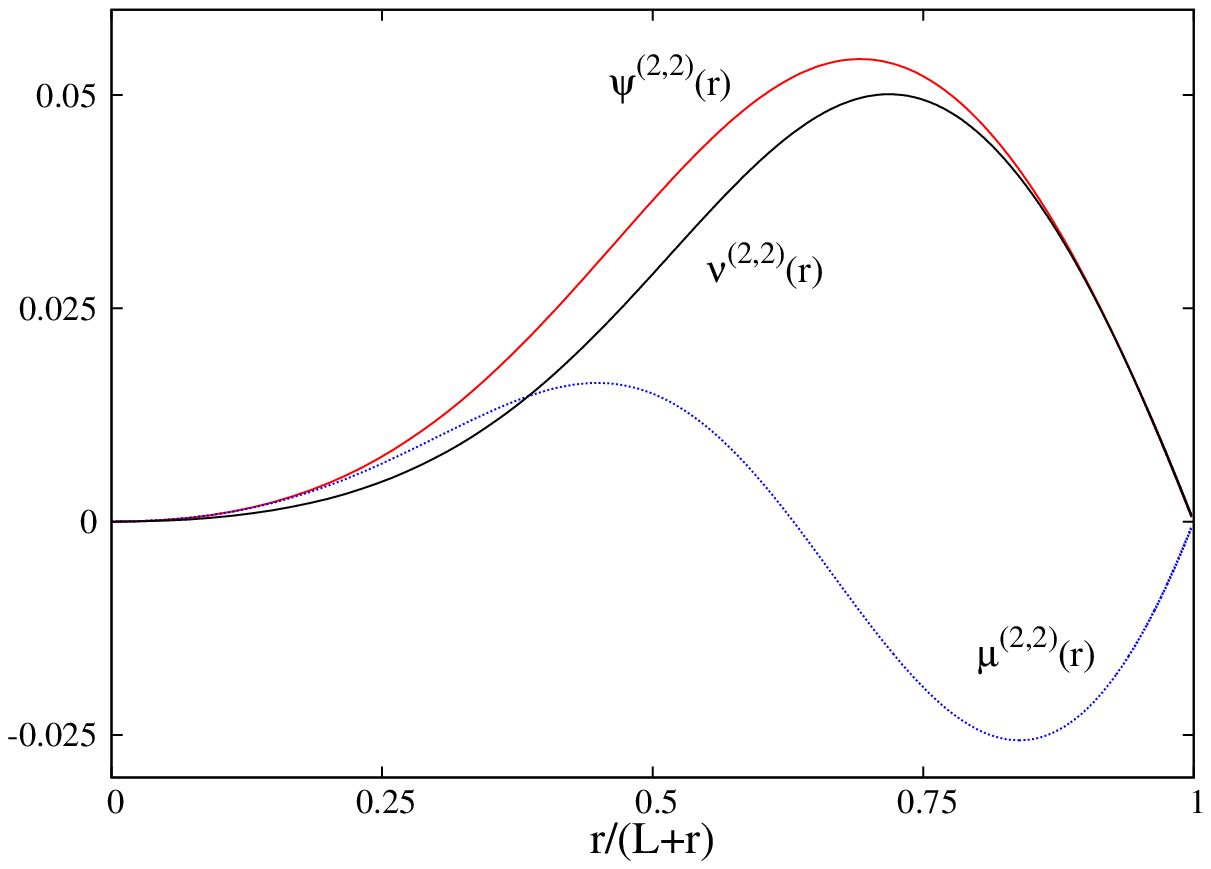}
\caption{Radial functions defining the perturbed metric (\ref{metrica}), via (\ref{mf1}) and (\ref{nu2})-(\ref{expansion}).}
\label{perturb}
\end{center}
\end{figure} 

To summarise, the perturbed metric to $\mathcal{O}(\alpha^2)$ by the $AdS$ regular electric dipole is given by (\ref{metrica}), with the metric functions expanded as (\ref{mf1}),
with the zeroth order contributions given by (\ref{order0}), and the second order contributions given by (\ref{nu2})-(\ref{psi20}) and (\ref{mu22})-(\ref{psi22}). In Figure~\ref{perturb} we plot the radial functions that define the perturbed metric. As can be seen they are bounded and smooth. 
Thus, at this order the backreacted metric is smooth and we did not 
notice the existence of any pathology.
Moreover,
we have verified that
the spacetime is asymptotically $AdS$,
according to the definition in  
\cite{Ashtekar:1999jx}.
The conformal boundary metric is given by the Einstein static universe,
while, at this order, the mass of the spacetime computed according to
 both the prescriptions in 
\cite{Ashtekar:1999jx}
and in 
\cite{Balasubramanian:1999re}
is still given by the soliton mass $E_1$ in (\ref{Ei}).

\section{Conclusions}
\label{sec_5}
In this paper we have considered electrostatic in global $AdS$. 
We have shown that there are everywhere regular multiple moments 
for $\ell\ge 1$. Interestingly, all such regular electric multipoles 
fall-off towards the $AdS$ boundary as $1/r$ in terms of the areal radius $r$. 
This decay invalidates a Lichnerowicz-type argument for the absence of solitons 
in Einstein-Maxwell-$AdS$. In fact we provide evidence that such solitons exist, 
by computed the perturbations induced in the $AdS$ geometry by a regular electric dipole, 
and observing all metric functions remain smooth. Showing the existence of fully 
non-linear solutions seems to require a numerical approach which is currently being considered. 

Finally, let us comment on various possible generalizations of the solutions in this work. 
Firstly, at the test field level, one can consider the dual magnetic multipoles, which would possess very similar properties. 
Moreover, superposing electric and magnetic multipoles it is possible to find configurations with non-vanishing angular momentum. 
We thus anticipate that the backreaction of such electromagnetic fields can lead to spinning solitons in Einstein-Maxwell-$AdS$ theory. 
Secondly, since $AdS$ is conformally flat and four dimensional Maxwell's theory is conformally invariant, any solution of Maxwell's equation in $AdS$ can be mapped to a solution of Maxwell's equations in Minkowski spacetime. 
It would be interesting to interpret the latter solutions corresponding to the regular electric multipoles described in this paper. 


\section*{Acknowledgements}
C.H. and E.R. thank funding from the FCT-IF programme, the grants PTDC/FIS/116625/2010 and   NRHEP--295189-FP7-PEOPLE-2011-IRSES, as well as the CIDMA research unit strategic funding UID/MAT/04106/2013.

\bigskip

\appendix

{

 \begin{small}
 
 \end{small}


\begin{thebibliography}{99}

\bibitem{Maldacena:1997re}
  J.~M.~Maldacena,
  Int.\ J.\ Theor.\ Phys.\  {\bf 38} (1999) 1113
   [Adv.\ Theor.\ Math.\ Phys.\  {\bf 2} (1998) 231]
  [hep-th/9711200].
\bibitem{BoutalebJoutei:1979va}
  H.~Boutaleb-Joutei, A.~Chakrabarti and A.~Comtet,
  Phys.\ Rev.\ D {\bf 20} (1979) 1884.
\bibitem{Bjoraker:2000qd}
  J.~Bjoraker and Y.~Hosotani,
  Phys.\ Rev.\ D {\bf 62} (2000) 043513
  [hep-th/0002098].
\bibitem{Bjoraker:1999yd}
  J.~Bjoraker and Y.~Hosotani,
  Phys.\ Rev.\ Lett.\  {\bf 84} (2000) 1853
  [gr-qc/9906091].
\bibitem{Boucher:1983cv}
  W.~Boucher, G.~W.~Gibbons and G.~T.~Horowitz,
  Phys.\ Rev.\ D {\bf 30} (1984) 2447.
\bibitem{Shiromizu:2012hb}
  T.~Shiromizu, S.~Ohashi and R.~Suzuki,
  Phys.\ Rev.\ D {\bf 86} (2012) 064041
  [arXiv:1207.7250 [gr-qc]].
\bibitem{Horowitz:2014hja}
  G.~T.~Horowitz and J.~E.~Santos,
  arXiv:1408.5906 [gr-qc].
\bibitem{Ashtekar:1999jx}
  A.~Ashtekar and S.~Das,
  Class.\ Quant.\ Grav.\  {\bf 17} (2000) L17
  [hep-th/9911230].
\bibitem{Balasubramanian:1999re}
  V.~Balasubramanian and P.~Kraus,
  Commun.\ Math.\ Phys.\  {\bf 208} (1999) 413
  [hep-th/9902121].
        
 \end{thebibliography}
\end{document}